\documentclass[%
 reprint,
 amsmath,amssymb,
 aps,
floatfix,
]{revtex4-2}

\usepackage[utf8]{inputenc}% Allow UTF-8 characters
\usepackage{CJKutf8}% CJK characters for Chinese name
\usepackage{graphicx}% Include figure files
\usepackage{dcolumn}% Align table columns on decimal point
\usepackage{bm}% bold math
\usepackage{booktabs}% For professional tables (\toprule, \midrule, \bottomrule)
\usepackage{array}
\usepackage{tabularx}
\usepackage{hyperref}% add hypertext capabilities
\usepackage{xurl}
\begin{document}
\title{Sparse M\"untz--Sz\'asz Recovery for Boundary-Anchored Velocity Profiles:\\ A Short-Record Roughness Diagnostic in Turbulence}% Force line breaks with \\

\newcommand{\CJKname}[1]{\begin{CJK}{UTF8}{gbsn}#1\end{CJK}}
\author{D Yang Eng (\CJKname{黄鼎扬})}
\email{dyang.eng@tum.de}

\affiliation{%
 Technical University of Munich, Arcisstraße 21, 80333 München, Germany\\
 }
 \affiliation{%
 Singapore Institute of Technology, 11 New Punggol Rd, Singapore 828616
 }

\date{\today}% It is always \today, today,

\newtheorem{theorem}{Theorem}
\newtheorem{definition}{Definition}
\newtheorem{remark}{Remark}
\newcommand{\kpop}{\kappa_{\mathrm{pop}}}

% \title{\textbf{Geometric Sharpness and Local Regularity in 3D Turbulence:\\ A Rigorous Empirical Diagnostic via M\"untz-Sz\'asz Sparse Recovery}}
% \author{D Yang Eng (黄鼎扬)}
% \date{\today}

% \maketitle

\begin{abstract}
We present a sparse convex-relaxation framework for estimating effective local scaling exponents from short boundary-anchored velocity-increment profiles ($N\approx40$). The detector solves an $\ell_1$-regularized regression in a mixed M\"untz--Sz\'asz/Jacobi dictionary and is interpreted throughout as a finite-scale, directional roughness diagnostic rather than a pointwise H\"older exponent. On isotropic datasets from the Johns Hopkins Turbulence Database, an internal subsampling benchmark against $N=200$ detector labels gives $F_1\approx0.93$ across nine unweighted reruns, and a balanced synthetic control gives balanced accuracy $0.928$ at $N=40$, indicating useful short-record self-consistency without constituting an external calibration. Across $Re_\lambda\approx433$--$1300$, the fixed-window sharp fraction remains of order $30$--$50\%$, but a scale-normalized control does not isolate a clean Reynolds-number trend. The recovered $\hat{\alpha}$ is only weakly associated with dissipation, whereas higher vorticity is consistently associated with smaller detected roughness exponents in conditioned samples. Directional controls on 60 high-vorticity centers further show a positive vorticity-aligned contrast $\Pi_\alpha$ (mean $0.093$, bootstrap 95\% CI $[0.028,0.158]$), stronger on the true vorticity axis than on fake axes, together with a statistically detectable low-order quadrupolar component in a joint Legendre fit. A seeded scale-transfer scan shows positive $\Pi_\alpha$ at both the smallest and largest tested radii, supporting finite-range persistence without a strong theorem-level nonlocal claim. The method is therefore best viewed as a finite-scale geometric diagnostic complementary to energetic observables, capable of resolving directional structure and low-order anisotropic organization in high-vorticity regions.
\end{abstract}
\maketitle

% ============================================================
\section{Introduction: Beyond Global Smoothness}
% ============================================================
This paper studies a local inverse problem: given a short boundary-anchored velocity-increment profile, infer whether its leading finite-scale behavior is better represented by a rough fractional power law or by a smooth polynomial background. We formulate that problem as sparse recovery in a mixed M\"untz--Sz\'asz/Jacobi dictionary and evaluate the resulting detector primarily as a reusable short-record diagnostic on canonical JHTDB isotropic datasets. The emphasis is methodological rather than proclamatory: we assess reproducibility, short-record self-consistency, and complementarity to standard energetic observables, while keeping the interpretation of $\hat{\alpha}$ deliberately finite-scale and directional.

\subsection{The intermittency problem}
Kolmogorov’s 1941 theory (K41) serves as the bedrock for understanding small-scale turbulence, providing the foundational approximation that velocity increments scale as $\delta u(\ell) \sim \ell^{1/3}$, which corresponds to a homogeneous Hölder exponent $h = 1/3$.\cite{frisch1995,sreenivasan1997}%
Yet, decades of experimental and numerical evidence reveal systematic deviations from this mean-field prediction.\cite{sreenivasan1997,wang1996,ishihara2009,kaneda2003}%
These departures, most visible in high-order statistics, point to the existence of intense, small-scale intermittency.\cite{sreenivasan1997,meneveau1991}%
While the 1962 refined similarity hypothesis (K62) attempted to bridge this gap through a lognormal model of dissipation fluctuations,\cite{kolmogorov1962,frisch1995}%
it is the multifractal formalism of Parisi and Frisch\cite{parisi1985} that offers a more complete statistical description.%
By introducing a continuous spectrum of local Hölder exponents $\alpha$ and a singularity spectrum $f(\alpha)$, this framework posits that the most extreme dissipative events occupy a set of vanishingly small measure, yet dominate the overall energy cascade.\cite{parisi1985,meneveau1991}

Despite the elegance of the multifractal model, verifying its predictions remains a persistent challenge.\cite{frisch1995,sreenivasan1997,benzi2009,dubrulle2019,dubrulle2022}%
Most current evidence for Reynolds-dependent intermittency amplification, such as that proposed by the log-Poisson cascade of She and L\'ev\^eque,\cite{she1994}%
is fundamentally indirect.%
We rely on global scaling exponents $\zeta_p$ calculated from massive datasets, rather than observing the local regularity of individual structures.\cite{sreenivasan1997,wang1996,ishihara2009}%
This reliance on global averages effectively masks the local physics.%
To bridge the gap between phenomenological theory and the Navier--Stokes equations, we require a diagnostic capable of probing effective local scaling around individual intense events. Critically, such a tool must function under the sparse-data constraints common in experimental settings, where traditional wavelet-transform modulus maxima (WTMM) or structure-function methods often fail to provide reliable local estimates.\cite{muzy1991,arneodo2002,jaffard2016}
\subsection{Limitations of classical diagnostics}

Classical structure-function analysis estimates anomalous scaling exponents $\zeta_p$ from relations of the form
$\langle |\delta u(\ell)|^p \rangle \sim \ell^{\zeta_p}$, providing a global characterization of intermittency but no per-event classification or spatial localization.\cite{frisch1995,sreenivasan1997,wang1996}
Reliable log--log regression requires signals spanning many decades of scale and typically several hundred to a thousand samples per record, especially at high orders where statistical convergence is slow.\cite{sreenivasan1997,ishihara2009,kaneda2003}
As a result, structure functions are poorly suited for the sparse or subgrid regimes common in experimental measurements and coarse DNS.

Wavelet-based multifractal formalisms, including the wavelet-transform modulus maxima (WTMM) method and later leader or $p$-leader approaches, offer local regularity estimates by tracking wavelet modulus maxima lines or aggregating local suprema of wavelet coefficients across scales.\cite{muzy1991,arneodo2002,jaffard2016}
These techniques have been successfully applied to turbulent fields and a wide range of fractal signals, but they assume access to long signals spanning many octaves of scale and are sensitive to boundary effects, particularly when the singularity of interest lies at a domain endpoint.
In practice, stable estimation of the local Hölder exponent with WTMM or $p$-leaders typically requires $\mathcal{O}(10^3)$ samples and interior singularities.\cite{muzy1991,arneodo2002,jaffard2016}

In many practical situations, for example, localized DNS probes, hot-wire measurements near intense structures, or radial velocity increment profiles of the form $f(r) = |u(x_0 + r \hat{n}) - u(x_0)|$ terminating at $r = 0$, the singularity of interest is located at a boundary and only $N \sim 10^1$--$10^2$ samples can be acquired. In this regime we interpret $\alpha$ as an effective scaling exponent over the finite sampled band rather than a strict pointwise H\"older exponent.
In this sparse and boundary-dominated regime, standard structure-function and WTMM-based methods become unreliable or inapplicable, motivating the development of diagnostics tailored to short radial profiles and boundary singularities.
Recent work on local Hölder exponents in turbulent vector fields has demonstrated the physical relevance of pointwise regularity diagnostics when full 3D DNS fields are available,\cite{meneveau1991,nguyen2019}
but an equivalent tool for sparse, one-dimensional measurements is still lacking.

\subsection{Geometric approaches to turbulence}

Beyond energetic descriptions based on the cascade, some recent work has emphasized \emph{geometric and structural} aspects of turbulent flows~\cite{sevilla2026}. We cite that perspective here only as motivation. The present manuscript does not test a full geometric-dynamics theory; rather, it introduces a sparse local diagnostic that may capture aspects of short-profile geometry not summarized by dissipation alone.

\subsection{Compressed sensing and geometric separation}

The detection problem considered here is naturally cast as a sparse approximation task in a structured dictionary.
Compressed sensing and sparse recovery theory have established that $\ell_1$-regularized estimators such as the Lasso can recover sparse signals from strongly undersampled linear measurements, provided the sensing matrix satisfies suitable incoherence or restricted isometry conditions.\cite{donoho2006,candes2011,bickel2009}
In the classical setting, dictionaries are often designed to have small mutual coherence or to satisfy a restricted isometry property, which ensures that different atoms are sufficiently separated and that convex relaxation accurately identifies the true support.\cite{donoho2006,candes2011}
Turbulent velocity profiles, however, are more naturally represented in highly redundant bases that mix singular and smooth components, and the resulting dictionaries are far from incoherent in the usual sense.

To capture the geometry of point singularities in a radial profile $f(r) = |u(x_0 + r \hat{n}) - u(x_0)|$, we construct a dictionary that explicitly splits a singular and a smooth subspace.
The singular part comprises fractional power functions $r^\alpha$ drawn from a Müntz--Szász sequence, which is known to be dense in $C([0,1])$ under mild divergence conditions on the exponent sequence.\cite{eng2026}
The smooth part is spanned by low-degree Jacobi polynomials, which approximate the regular background flow.
This design reflects the physical decomposition of a turbulent velocity increment into a localized singular core and a smoother surrounding profile, and follows the geometric separation framework developed in Ref.~\cite{eng2026} for ultra-coherent dictionaries.

From the standpoint of standard sparse recovery theory, the resulting dictionary is highly coherent: nearby exponents $\alpha$ produce very similar fractional atoms, and the singular atoms can have substantial overlap with the low-degree polynomial subspace.\cite{eng2026}
In such regimes, textbook bounds based on mutual coherence or restricted isometries predict large sample requirements and do not distinguish between adversarial and physically structured signals.\cite{donoho2006,candes2011}
The Müntz--Szász geometric separation theory introduces a Singular Separation Condition (SSC) and a population gap parameter $\kappa_{\mathrm{pop}}(\alpha)$ that quantify the distance between the singular and smooth subspaces.\cite{eng2026} In the present paper, we use that geometry only as motivation for the dictionary design and as a qualitative descriptor of singular-vs.-smooth separation. We do not make quantitative sample-complexity comparisons for the unweighted discrete detector used in the reported experiments.
\subsection{Contributions}

We emphasize a crucial distinction from multifractal formalisms: our method estimates an \emph{effective} scaling exponent over a finite radial window ($r \in [0, r_{\max}]$) in a single random direction. This makes $\hat{\alpha}$ a \textbf{finite-scale, directional roughness indicator} rather than a true pointwise H\"older exponent. The distinction matters because:
\begin{itemize}
\item The $r^\alpha$ model is a local approximation, not a claim about pointwise regularity;
\item Finite $r_{\max} \approx 70\,\eta_K$ means we probe scaling over a bounded range, not asymptotic behavior;
\item Single-direction sampling captures directional roughness, not isotropic regularity.
\end{itemize}

Principal contributions:
\begin{itemize}
\item A sparse-recovery framework based on a fractional-polynomial dictionary for estimating an effective exponent $\alpha$ from $N \approx 40$ boundary-anchored samples. Throughout, $\hat{\alpha}$ is interpreted as a \emph{finite-scale, directional roughness indicator} and sparse geometric observable rather than a pointwise H\"older exponent.
\item An internal JHTDB subsampling benchmark showing that, relative to $N{=}200$ detector labels on the same profiles, the $N{=}40$ sharp/smooth classifier attains high but run-dependent self-consistency; across nine saved unweighted reruns, $F_1$ spans $0.84$ to $1.00$ with mean $0.930$, and a recent unweighted run gives $0.927$. This benchmark is intended as a reproducible short-record utility check for the implemented detector, not as an external accuracy claim.
\item An empirical Reynolds-number study over three JHTDB isotropic datasets showing conditional sharp fractions in the range $30$\%--$50$\%, with overlapping confidence intervals and strong sensitivity to the probing window.
\item A dissipation comparison indicating that detected roughness is only weakly associated with local $\epsilon$, supporting the interpretation that $\hat{\alpha}$ carries information about profile geometry not exhausted by energetic amplitude alone, together with a short Lagrangian study showing rapid loss of persistence under a direction-resampled protocol.
\item A finite-scale directional observable family built from the same short-profile estimator, including the vorticity-aligned contrast $\Pi_\alpha=\hat{\alpha}_{\parallel}-\hat{\alpha}_{\perp}$ and the anisotropy summaries $\Delta_\alpha$ and $\mathrm{Var}(\hat{\alpha})$, designed to probe local geometric organization in a form that can be reused in later theoretical or numerical work.
\item A reproducible, explicitly model-dependent diagnostic for sparse local-profile analysis, intended for use alongside standard energetic observables and full-field multifractal methods rather than as a replacement for them.
\end{itemize}
\noindent
\section{Methodology}
% ============================================================

\subsection{Data Source: Johns Hopkins Turbulence Database}
We analyze the \texttt{isotropic1024coarse} dataset from the Johns Hopkins Turbulence Database (JHTDB),\cite{jhtdb}
a direct numerical simulation of forced isotropic turbulence at Taylor-scale Reynolds number $Re_\lambda \approx 433$,
stored on a $1024^3$ periodic grid with spatial resolution $\Delta x = 2\pi / 1024 \approx 0.00614$.
Throughout, quoted $Re_\lambda$ values are the nominal dataset values used in the repository experiment configurations; they are not recomputed from raw fields within this manuscript.
The Kolmogorov microscale for this dataset is $\eta_K \approx 0.00287$ (computed from $\nu = 0.000185$ and $\langle \epsilon \rangle = 0.0928$),\cite{jhtdb}
so the DNS grid resolves approximately $2.2\,\eta_K$.

\paragraph{Profile extraction.} For each candidate point $\mathbf{x}_0$, we extract a one-dimensional radial velocity increment profile:
\begin{equation}\label{eq:profile}
    f(r) = |\mathbf{u}(\mathbf{x}_0 + r\,\hat{\mathbf{n}}) - \mathbf{u}(\mathbf{x}_0)|, \qquad r \in [0, r_{\max}],
\end{equation}
where $\hat{\mathbf{n}}$ is a random unit direction and $r_{\max} = 0.2$ for the baseline JHTDB experiments. This profile is sampled at $N$ equispaced points along the ray. The default implementation uses the magnitude increment rather than the longitudinal increment $\mathbf{u} \cdot \hat{\mathbf{n}}$; a separate control experiment compares the two choices. Candidate points are selected by thresholding the vorticity magnitude returned by JHTDB gradient queries. In the current codebase this threshold is a user-specified numerical cutoff (typically 5, 10, or 12 in the queried units), not a standardized multiple of $\sigma_\omega$.

\paragraph{Preprocessing and normalization.} The radial coordinate is rescaled to the unit interval $[0,1]$ before dictionary evaluation, with $f(0) = 0$ by construction. The current implementation does \emph{not} divide profile amplitudes by a global $u_{\mathrm{rms}}$ or any other dataset-wide scale. Instead, the detector operates on raw increment magnitudes and uses fixed hyperparameters $\sigma = 0.01$ and $|\hat{c}_j| > 10^{-5}$. These settings work reasonably in the reported experiments, but they should be read as empirical choices rather than as a demonstrated scale-invariant normalization.

\begin{table}[ht]
\centering
\caption{Sample size $N$ and corresponding physical resolution for $r_{\max}=0.2$ ($\approx 70\,\eta_K$) with $\eta_K \approx 0.00287$.}
\label{tab:resolution}
\begin{tabular}{ccc}
\hline
$N$ & $\mathrm{d}r$ (physical) & $\mathrm{d}r/\eta_K$ (resolution) \\
\hline
40  & $5.0 \times 10^{-3}$ & 1.74 (coarser than $\eta_K$) \\
60  & $3.3 \times 10^{-3}$ & 1.16 ($\approx \eta_K$) \\
80  & $2.5 \times 10^{-3}$ & 0.87 (sub-Kolmogorov) \\
120 & $1.7 \times 10^{-3}$ & 0.58 (well-resolved) \\
200 & $1.0 \times 10^{-3}$ & 0.35 (oversampled) \\
\hline
\end{tabular}
\end{table}

Table~\ref{tab:resolution} gives the approximate physical spacing associated with different $N$. We use it only as a scale guide. The sample-complexity experiment discussed later is defined relative to detector outputs at $N=200$, so it should not be interpreted as a direct physical proof of a sharp Kolmogorov-scale resolution threshold.

\subsection{Signal Model}
We model the local velocity increment as a superposition of a singular and a smooth component:
\begin{equation}\label{eq:signal_model}
    f(r) = c_\alpha\, r^{\alpha} + \sum_{k=0}^{K} c_k\, \phi_k(r) + \varepsilon(r),
\end{equation}
where $r^\alpha$ is the singular atom with effective scaling exponent $\alpha \in (0,1)$ over the sampled band, $\{\phi_k\}_{k=0}^K$ are low-degree Jacobi polynomials (the ``smooth'' subspace), and $\varepsilon$ represents aggregate residuals, including measurement noise and model mismatch. Equation~\eqref{eq:signal_model} is an approximation ansatz used for sparse fitting; the current experiments do not prove that DNS profiles satisfy this decomposition exactly or that a uniform bound such as $\|\varepsilon\|_\infty \leq \sigma$ holds profile by profile.

\subsection{Dictionary Construction}
The combined dictionary $\mathbf{D} = [\boldsymbol{\Psi} \mid \boldsymbol{\Phi}] \in \mathbb{R}^{N \times D}$ consists of:

\begin{itemize}
    \item \textbf{Singular atoms} $\boldsymbol{\Psi}$: M\"untz basis functions $\psi_j(r) = r^{\alpha_j} / \|r^{\alpha_j}\|_{L^2(\mu)}$, evaluated on a chosen exponent grid $\{\alpha_j\}_{j=1}^{n_\alpha} \subset (0,1)$. The $L^2(\mu)$ norm is computed under the Jacobi measure $d\mu = r^\beta(1-r)^b\,dr$ with $\beta = 0.55$, $b = 0$.
    \begin{equation}
        \|r^\alpha\|_{L^2(\mu)}^2 = \int_0^1 r^{2\alpha + \beta}(1-r)^b\,dr = B(2\alpha + \beta + 1, b + 1).
    \end{equation}

    \item \textbf{Smooth atoms} $\boldsymbol{\Phi}$: Jacobi polynomials $\phi_k(r) = P_k^{(b,\beta)}(2r-1) / \|P_k\|$, $k = 0, \ldots, K$ with $K = 4$, giving $K+1=5$ smooth atoms under the same measure $\mu$.
\end{itemize}

The detector class accepts several grids. In the present repository, dictionary-geometry calculations and some synthetic studies use $n_\alpha=37$ exponents on $[0.05,0.95]$, while several JHTDB experiments use $n_\alpha=21$ exponents on $[0.1,0.9]$. Accordingly, the total dictionary size is either $D=42$ or $D=26$ in the current experiments. We therefore interpret each reported result with the grid used by the generating script rather than claiming that all figures come from a single universal discretization.

\begin{table}[t]
\centering
\caption{Detector grids used by the manuscript figures. ``Fine'' denotes 37 exponents on $[0.05,0.95]$ ($D=42$), while ``standard'' denotes 21 exponents on $[0.10,0.90]$ ($D=26$). Figure~\ref{fig:dict_geometry} reports analytic weighted-model geometry and is not tied to a single discrete detector grid.}
\label{tab:grid_usage}
\begin{tabular}{ll}
\hline
Figure or result block & Detector grid \\
\hline
Figs.~\ref{fig:synth_scatter} and \ref{fig:baseline} & fine (37-point) \\
Fig.~\ref{fig:phase_transition} & standard (21-point) \\
Figs.~\ref{fig:dissipation}, \ref{fig:lagrangian}, and \ref{fig:reynolds} & fine (37-point) \\
Text-only auxiliary controls (multi-snapshot, topology) & standard (21-point) \\
Supplementary Fig.~\ref{fig:supp_s1} & fine (37-point) \\
\hline
\end{tabular}
\end{table}

\begin{remark}[Grid truncation]
In the current codebase, all reported experiments restrict $\alpha$ to either $[0.05,0.95]$ or $[0.1,0.9]$. Exponents near $1$ or above are therefore not resolved explicitly. Profiles with near-linear or super-linear scaling may be absorbed into the largest available smooth-looking bin, so the method should not be interpreted as a viscous-range regularity estimator.
\end{remark}

\begin{figure}[t]
  \centering
  \includegraphics[width=\columnwidth]{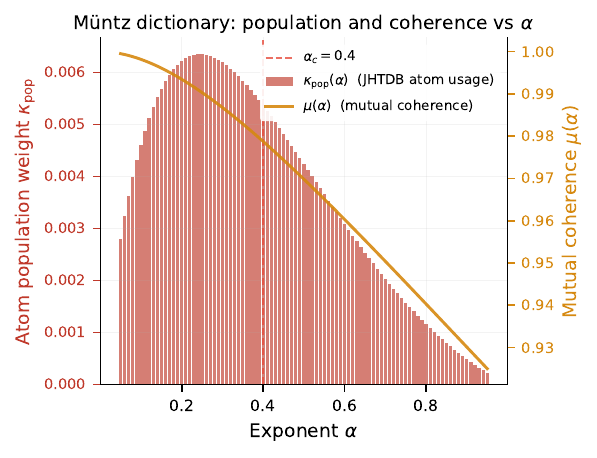}
  \caption{Analytic dictionary geometry. Bars show the computed
    $\kappa_{\mathrm{pop}}(\alpha)$ curve, and the line shows the
    singular-vs.-polynomial cross-coherence proxy
    $\mu_{\mathrm{cross}}(\alpha)$. These are continuous weighted-model
    diagnostics; they are not empirical detection frequencies.}
  \label{fig:dict_geometry}
\end{figure}

There is a basic geometry mismatch between theory and implementation. The analytic quantities $\kpop(\alpha)$ and the cross-coherence proxy discussed below are computed under a continuous weighted inner product, whereas the detector is applied to discrete equispaced samples and, by default, uses an unweighted Lasso objective. We therefore use the geometric quantities as qualitative diagnostics, not as exact predictors of finite-sample performance. Figure~\ref{fig:dict_geometry} reports the analytic curves actually computed in the repository; it does not show an empirical detection-frequency histogram.

\subsection{LASSO Detection}
Detection proceeds by solving the $\ell_1$-regularized least squares problem:
\begin{equation}\label{eq:lasso}
    \hat{\mathbf{c}} = \arg\min_{\mathbf{c}} \frac{1}{2N}\|\mathbf{f} - \mathbf{D}\mathbf{c}\|_2^2 + \lambda_N \|\mathbf{c}\|_1,
\end{equation}
where $\mathbf{f} = (f(r_1), \ldots, f(r_N))^\top$. The implementation sets
\begin{equation}\label{eq:lambda}
    \lambda_N = 4\sigma\sqrt{\frac{\log D + \log(1/\delta)}{N}},
\end{equation}
with $\sigma = 0.01$ and $\delta = 0.05$. This choice is motivated by \cite{eng2026}, but in the current code it is used as a heuristic regularization rule rather than as a verified finite-sample guarantee. By default, the implementation sets \texttt{use\_quadrature\_weights=False}, so Eq.~\eqref{eq:lasso} is solved on unweighted samples even though the dictionary is motivated by a weighted analytic geometry. A preliminary quadrature-weighted variant was explored during revision, but under the present boundary-anchored JHTDB protocol it materially shifted the detector operating point and did not preserve the benchmark behavior of the reported pipeline. We therefore do not present that weighted variant as a drop-in replacement in this manuscript. A singularity at exponent $\alpha_j$ is declared detected if $|\hat{c}_j| > 10^{-5}$.

\paragraph{Handling of null detections.} When no singular atom exceeds the threshold $\tau = 10^{-5}$ (i.e., $\|\hat{\mathbf{c}}_{\text{sing}}\|_\infty \leq \tau$), the profile is classified as \textbf{smooth} with $\hat{\alpha}$ set to $\mathrm{NaN}$ and excluded from $\alpha$-distribution summaries. In the current DNS experiments, the fraction of requested profiles without a reported representative $\alpha$ is typically several percent and can approach $10\%$, depending on dataset and protocol. These cases are retained in binary counts only when the corresponding script does so explicitly.

\subsection{Exponent Refinement}
After grid detection, the detected exponent is refined by continuous optimization. For the dominant detected atom $\alpha_{\text{grid}}$ (the active atom with largest $|\hat{c}|$), we solve:
\begin{equation}\label{eq:refine}
    \hat{\alpha} = \arg\min_{\alpha \in [\alpha_{\text{grid}} - w,\, \alpha_{\text{grid}} + w]} \|f - \Pi_\alpha f\|_2^2,
\end{equation}
where $\Pi_\alpha$ denotes the orthogonal projection onto $\text{span}(r^\alpha, \phi_0, \ldots, \phi_K)$ and $w = 0.05$ is the refinement window. This is solved via Brent's bounded scalar optimization \cite{scipy}.

\subsection{Theory-to-Implementation Scope}
Reference \cite{eng2026} provides a geometric framework for reasoning about separation between singular and smooth dictionary components. In the present manuscript, that theory is used only as background motivation for the dictionary construction, the SSC terminology, and the analytic separation curves plotted later. We do not claim a verified theorem-to-implementation correspondence for the reported detector, and we do not compare the experiments against formal sample-complexity bounds. A revision-stage quadrature-weighted ablation was informative precisely because it changed the empirical operating point substantially in the current boundary-anchored setting; accordingly, we treat weighted-detector revalidation as future work rather than as something established here.

\subsection{Geometric Separation}
The population separation $\kpop(\alpha)$ measures the distance from a singular atom to the smooth polynomial subspace:
\begin{equation}\label{eq:kappa}
    \kpop(\alpha) = \inf_{p \in \mathcal{P}_K} \|r^\alpha - p\|_{L^2(\mu)} = \sqrt{\|r^\alpha\|^2 - \sum_{k=0}^{K} |\langle r^\alpha, \phi_k\rangle|^2}.
\end{equation}
This quantity governs the fundamental resolvability of a singularity: when $\kpop \to 0$, the singular atom becomes indistinguishable from the smooth subspace, and detection requires exponentially more samples. For our dictionary parameters ($K=4$, $\beta = 0.55$), we find $\kpop(0.4) \approx 0.005$, placing the detection problem firmly in the ultra-coherent regime.

\begin{remark}[Coherence metrics: $\mu$ vs.~$\kpop$]
The analytic geometry reports the continuous weighted cross-coherence proxy
\[
\mu_{\mathrm{cross}}(\alpha) = \max_{0 \le k \le K} |\langle \psi_\alpha, \phi_k \rangle_{L^2(\mu)}|,
\]
that is, coherence between singular atoms and the polynomial subspace. In
addition, the current repository now includes a direct discrete unweighted
diagnostic of the actual implementation design matrix on equispaced samples.
Across both detector grids and $N \in \{40,80,200\}$, the full pairwise
discrete mutual coherence satisfies $\mu_{\mathrm{global}} \approx
0.99990$--$0.99996$, while the singular-vs.-polynomial discrete cross
coherence already lies in the range $\mu_{\mathrm{cross}} \approx
0.984$--$0.996$. The smallest singular values are at machine precision and
the computed condition numbers of both $\mathbf{D}$ and
$\mathbf{D}^{\!\top}\mathbf{D}$ are numerically infinite in this diagnostic.
Thus the implemented unweighted detector also operates in an ultra-coherent,
nearly rank-degenerate discrete regime. These diagnostics strengthen the
cautionary empirical reading of the Lasso separator; they do not, by
themselves, supply a recovery theorem for the reported pipeline.
\end{remark}

\subsection{Classification Criterion}
A profile is classified as \textbf{intermittent} (sharp) if the representative exponent satisfies $\hat{\alpha} < \alpha_{\text{crit}} = 0.4$, and \textbf{smooth} otherwise. This threshold defines a practical intermittent class in the finite-band sense, and we report $f(\alpha < 0.4)$ throughout.

\subsection{Exponent Estimation}\label{sec:alpha_est}

After LASSO detection, the representative exponent is selected from the detected support. The implementation uses the \emph{dominant coefficient} criterion: the atom with largest absolute coefficient magnitude:
\begin{equation}\label{eq:alpha_est_corrected}
    \hat{\alpha} = \alpha_{j^*}, \quad j^* = \arg\max_j \{|\hat{c}_j| : |\hat{c}_j| > \tau\}.
\end{equation}
This differs from the ``minimum detected value'' (most singular) criterion. The dominant-coefficient selection is more stable numerically but may bias estimates toward mid-range exponents when multiple atoms activate. The ``most\_singular\_alpha'' (minimum) is computed for diagnostic purposes but not used as the primary estimate.

Our detection pipeline applies empirical thresholds for residual scale and coefficient magnitude ($\sigma = 0.01$, $|\hat{c}_j| > 10^{-5}$). Because the current implementation does not normalize amplitudes by $u_{\mathrm{rms}}$, these thresholds should be understood as practical settings rather than as a proven scale-invariant decision rule.

\section{Validation \& Robustness}

\subsection{Validation on Controlled Signals}

Before applying the detector to turbulence, we examine several controlled checks. They are useful for numerical sanity, but they do not amount to a full external calibration.

\paragraph{Fractional Brownian motion (fBm).} The repository includes a covariance-based fBm stress test using boundary increments $|B_H(r)-B_H(0)|$ on $r \in [0,1]$. This is useful as a qualitative rough-profile check, but the current outputs remain too biased and variable for a quantitative calibration claim. We therefore do not use fBm as evidence of detector calibration in this manuscript.

\paragraph{Synthetic singularities.} The primary synthetic control now uses \texttt{run\_synthetic\_benchmark.py}, which applies the actual Sparse Lasso detector to a balanced family of 200 sharp and 200 smooth boundary-anchored profiles per sample size, with random low-degree polynomial backgrounds and additive Gaussian noise. At the operating point $N=40$, $\sigma=0.01$, this balanced benchmark yields accuracy $=\,0.9275$, balanced accuracy $=\,0.9275$, precision $=\,0.873$, recall $=\,1.000$, specificity $=\,0.855$, and $F_1=0.932$, with no null detections. At $N=80$ and $N=200$, recall remains $1.000$ but specificity decreases to $0.820$ and $0.790$, respectively, so the synthetic evidence is best read as a high-recall numerical sanity check with nontrivial false positives rather than as a calibration proof. A separate script, \texttt{run\_exp\_i\_extended\_baselines.py}, remains useful for method comparison against OMP, ElasticNet, and a structure-function baseline, but its synthetic stage is still restricted to sharp profiles only.

\paragraph{What these validate.} The controlled checks support internal numerical behavior of the sparse detector on simplified signals. They do \textbf{not} validate pointwise H\"older estimation in turbulence, and they do not provide an external ground-truth calibration for the JHTDB studies.

\subsection{Algorithmic Robustness}
% ============================================================
We use three auxiliary robustness probes.
First, the analytic dictionary calculations report $\kpop(\alpha)$ in the range
$2.2\times 10^{-4}$ to $6.3\times 10^{-3}$ over the tested $\alpha$ grid,
together with a large singular-vs.-polynomial cross-coherence proxy. These
numbers characterize the continuous weighted model, not the discrete unweighted
detector used in the DNS experiments. Complementing those curves, the new
discrete unweighted dictionary diagnostic computes the full pairwise mutual
coherence and conditioning of the actual equispaced design matrices used by
the implementation. Across both grids and $N \in \{40,80,200\}$, the full
discrete mutual coherence is $\mu_{\mathrm{global}} \approx
0.99990$--$0.99996$, the singular-vs.-polynomial cross coherence is already
$0.984$--$0.996$, and the smallest singular values are at machine precision,
so the reported detector is operating in an extremely ill-conditioned discrete
regime as well.

Second, the script \texttt{run\_exp\_c\_sample\_complexity.py} subsamples
high-vorticity JHTDB profiles from $N=200$ down to smaller $N$ and uses the
detector output at $N=200$ as the reference label for the same profile. In a
recent unweighted run, the overall sharp/smooth classifier has precision
$0.95$, recall $0.905$, and $F_1 = 0.927$ at $N=40$. Across the nine saved
unweighted reruns used in the manuscript summary, the corresponding $N=40$
$F_1$ value ranges from $0.84$ to $1.00$ with mean $0.930$. Here ``nine
reruns'' means the eight archived pre-transition root artifacts from March
1--21 plus the guarded post-revert run
\texttt{exp\_c\_sample\_complexity\_20260323\_002345.json}; the two March 23
transition artifacts generated during the temporary quadrature-weighting test
(\texttt{20260323\_000652} and \texttt{20260323\_000857}) are excluded. These
are self-consistency numbers under subsampling, not recovery against external
ground truth, and they should be read as run-dependent internal validation
rather than as a single fixed performance constant.

Third, an auxiliary $\lambda$-sweep on JHTDB profiles finds an empirical region
in which the detected support is relatively stable and the heuristic
$\lambda_N$ falls inside that region. We treat this as a descriptive robustness
check rather than as a proof of optimality in an ultra-coherent regime.
\begin{figure}[t]
  \centering
  \includegraphics[width=\columnwidth]{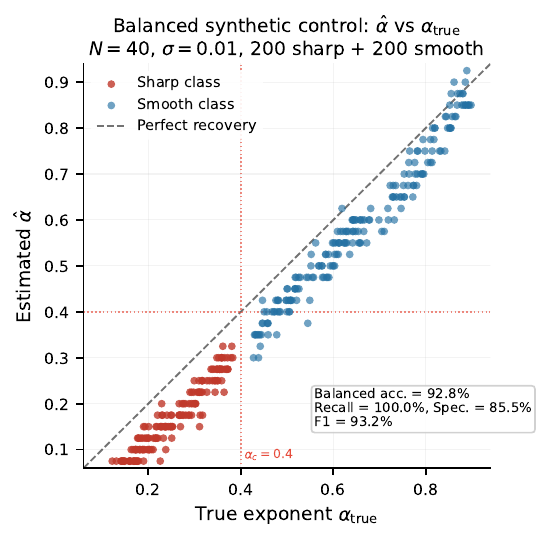}
  \caption{Balanced synthetic control: estimated $\hat\alpha$ vs.\ true
    $\alpha_{\mathrm{true}}$ for Sparse LASSO at $N{=}40$, $\sigma{=}0.01$,
    with 200 sharp and 200 smooth synthetic profiles. At this operating
    point, the detector attains accuracy $=\,0.9275$, balanced accuracy
    $=\,0.9275$, precision $=\,0.873$, recall $=\,1.000$, specificity
    $=\,0.855$, and $F_1=0.932$, with no null detections. The visible spread
    in the smooth class reflects nontrivial false positives, so this figure
    supports numerical sanity rather than perfect synthetic classification.}
  \label{fig:synth_scatter}
\end{figure}
\begin{figure}[t]
  \centering
  \includegraphics[width=\columnwidth]{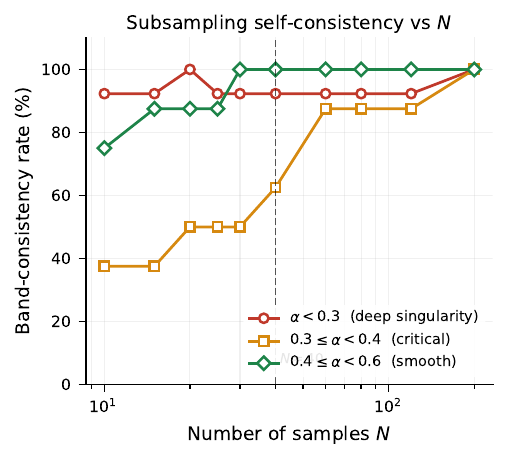}
  \caption{Subsampling study on JHTDB profiles. The panel plots detection
    rates for three reference-$\alpha$ bands as a function of $N$. In a
    recent unweighted run, the same experiment yields an overall sharp/smooth
    $F_1$ score of $0.927$ at $N{=}40$ when $N{=}200$ detector outputs on the
    same profiles are used as reference labels. Across the nine saved
    unweighted reruns used in the manuscript summary, the corresponding
    $N{=}40$ $F_1$ value spans $0.84$ to $1.00$; the two March 23 transition
    artifacts from the temporary quadrature-weighting test are excluded from
    that count. This figure should therefore be interpreted as an internal
    run-dependent self-consistency study. The vertical marker is a heuristic
    guide used in the plotting script.}
  \label{fig:phase_transition}
\end{figure}

\subsection{Comparative Performance}

The sparse M\"untz--Sz\'asz framework is designed for boundary-anchored
profiles where classical methods fail. WTMM requires interior singularities
accessible from both sides, while structure functions estimate bulk exponents
and require $\gg N$ samples for stable log--log fits. 

On the balanced synthetic control at $N=40$, Sparse LASSO achieves perfect
sharp-class recall but imperfect specificity, so its synthetic performance is
strong but not trivialized by the absence of smooth counterexamples. In a
separate restricted sharp-only method-comparison benchmark, LASSO and OMP
(both using the same dictionary) detect all implanted sharp profiles, while
ElasticNet and the structure-function baseline perform poorly (3.3\% and
0.0\% detection). On the 19 reference-sharp JHTDB profiles within the pinned
50-profile method-comparison artifact, LASSO detects 78.9\% of
reference-sharp profiles, OMP 63.2\%, ElasticNet 15.8\%, and the
structure-function baseline 0.0\%. These numbers indicate that sparse
dictionary methods are better matched to the present short-profile geometry
than the simple baselines tested here, but they should be read as a small-$n$
method comparison rather than as a universal ranking across all
local-regularity methods, because WTMM and wavelet-leader approaches are not
implemented on a common benchmark appropriate to their sampling assumptions.

\begin{figure}[t]
  \centering
  \includegraphics[width=\columnwidth]{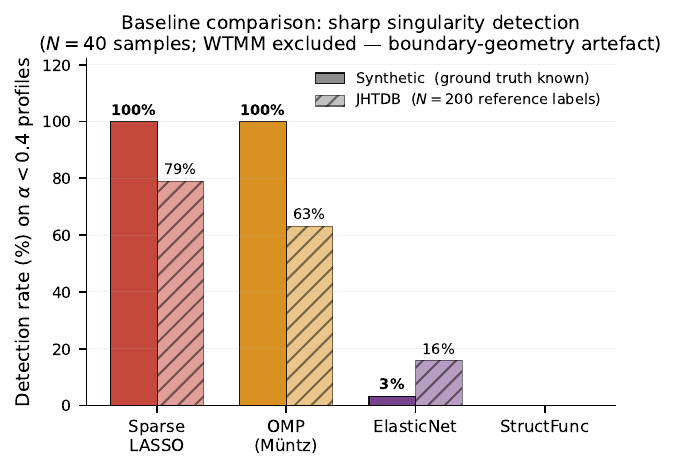}
  \caption{Baseline comparison at $N=40$. The synthetic panel in this figure
    is a restricted sharp-only method-comparison benchmark, whereas the
    balanced synthetic control appears separately in Fig.~\ref{fig:synth_scatter}.
    The JHTDB panel uses reference labels produced by the same detector at
    $N=200$ on the same profiles. Under those conventions, sparse methods
    outperform the simple ElasticNet and structure-function baselines tested
    here.}
  \label{fig:baseline}
\end{figure}

\subsection{Geometric Universality}

To probe sensitivity to sampling choices, we run auxiliary controls over time,
coherent-structure topology, direction, and detector hyperparameters.

Across five snapshots ($t \in \{0,2,4,6,8\}$), 1000 profile attempts yield
919 valid detections. The pooled sharp fraction is $35.36\%$ with a 95\% CI of
$[32.32\%, 38.52\%]$. This suggests that the intermittency level remains of the
same order across the sampled times, but the evidence is descriptive rather
than a formal test of temporal stationarity.

Using the $Q$--$R$ invariants of the velocity-gradient tensor to classify
high-vorticity points as tube-like ($Q>0$) or sheet-like ($Q<0$), the saved
topology experiment yields 102 tube points and 98 sheet points. The
corresponding sharp fractions are $48.96\%$ (tubes) and $37.78\%$ (sheets),
with mean detected exponents $0.453$ and $0.509$, respectively. These
differences are suggestive but remain conditional on the chosen topology
classifier and sampling rule.

Directional consistency is assessed by detecting $\alpha$ along the six
axis-aligned directions at each of 60 high-vorticity centers. The directional
standard deviation is similar for tubes and sheets
($\sigma_{\mathrm{dir}} \approx 0.240$ vs.\ $0.237$, $p=0.87$), indicating no
strong topology-dependent anisotropy in this particular probe.

\begin{figure}[t]
  \centering
  \includegraphics[width=0.48\columnwidth]{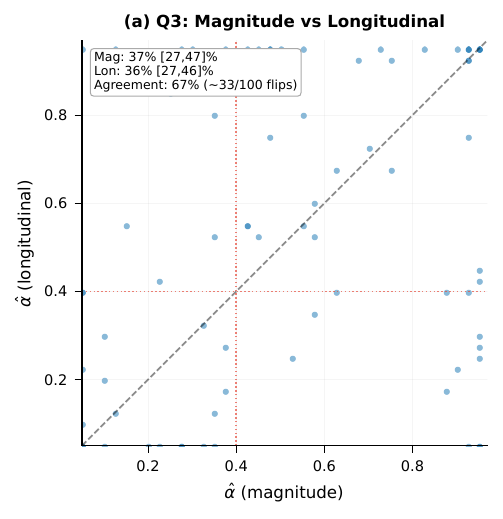}
  \includegraphics[width=0.48\columnwidth]{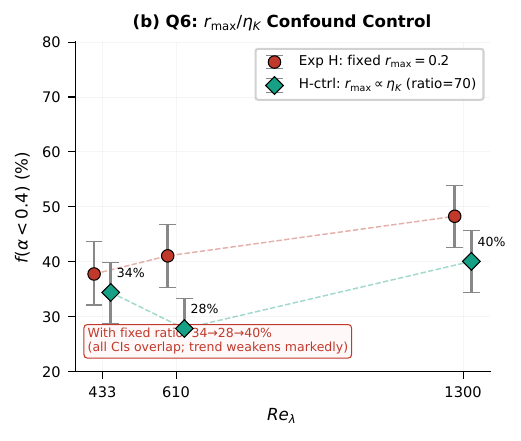}
  \includegraphics[width=0.48\columnwidth]{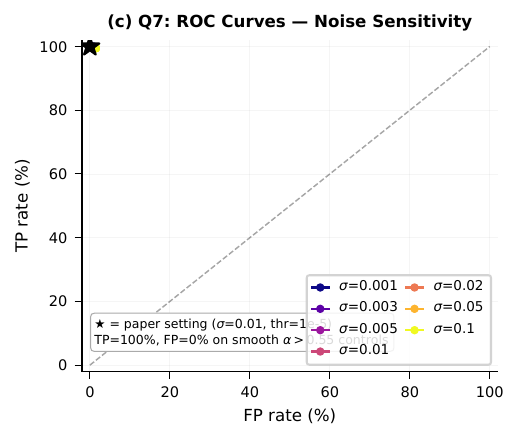}
  \includegraphics[width=0.48\columnwidth]{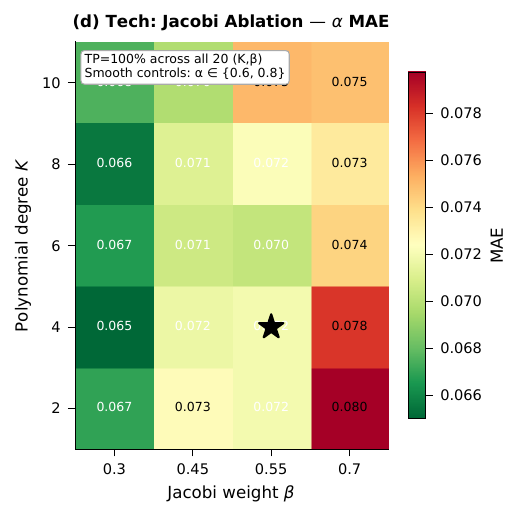}
  \caption{Targeted control experiments: (a) magnitude vs.\ longitudinal increments; (b) $r_{\max}/\eta_K$-normalized Reynolds control; (c) widened-margin sensitivity to $\sigma$ and coefficient threshold; (d) widened-margin Jacobi degree $K$ and weight $\beta$ ablation.}
  \label{fig:supp_s1}
\end{figure}

We also perform targeted control experiments to probe methodological choices.
Using identical JHTDB queries, magnitude and longitudinal increments yield
nearly identical \emph{marginal} detection rates at $N=40$ (37.0\% vs.\ 36.0\%)
with overlapping 95\% CIs and a negligible paired mean difference
$\langle \alpha_{\mathrm{mag}}-\alpha_{\mathrm{lon}}\rangle=-0.012$. However,
the paired sharp/smooth calls agree on only 67\% of the 100 requested
profiles, so roughly one-third flip class between increment definitions. We
therefore interpret increment type as having limited effect on aggregate
fractions but non-negligible effect on individual profile labels
(Fig.~\ref{fig:supp_s1}a). A
scale-normalized Reynolds control with $r_{\max}/\eta_K=70$ yields
$f(\alpha<0.4)=34.4\%, 27.8\%, 40.1\%$ across
$Re_\lambda=\{433,610,1300\}$ with overlapping CIs, showing that the monotonic
trend under fixed physical $r_{\max}$ is not robust to the scale window
(Fig.~\ref{fig:supp_s1}b). A direct vorticity-threshold sensitivity sweep on
\texttt{isotropic1024coarse}, using a shared candidate pool of 400 centers and
120 requested profiles per threshold, shows that the requested-profile sharp
fraction rises from $28.3\%$ at $|\omega|>3$ to $35.8\%$ at $|\omega|>12$,
while the valid-only fraction rises from $30.6\%$ to $37.7\%$. Across this
sweep, the Spearman correlation between $\hat{\alpha}$ and $|\omega|$ remains
negative for all tested thresholds, ranging from $\rho=-0.330$ to
$\rho=-0.453$ with $p<4\times10^{-4}$. Thus, within the conditioned
high-vorticity sample, larger vorticity magnitude is consistently associated
with smaller detected roughness exponents. Sensitivity of the detector
classification threshold is also non-negligible: using the valid detections
from the baseline $|\omega|>10$ slice ($n=112$), a post-processing sweep over
$\alpha_{\mathrm{crit}}\in[0.30,0.50]$ changes the reported sharp fraction
from $17.9\%$ at $\alpha_{\mathrm{crit}}=0.30$ to $47.3\%$ at
$\alpha_{\mathrm{crit}}=0.50$, with the manuscript operating point
$\alpha_{\mathrm{crit}}=0.40$ giving $35.7\%$ (95\% bootstrap CI
$[26.8\%,44.6\%]$). We therefore interpret absolute sharp fractions as
threshold-dependent summaries of a low-$\alpha$ tail rather than as
detector-independent population constants. Sensitivity of the detector
hyperparameters is also assessed on two auxiliary widened-margin synthetic
controls. In the $\sigma$/$|\hat{c}_j|$ sweep, the smooth class is sampled from
$\alpha\in[0.55,0.90]$, so the operating point
($\sigma=0.01$, $|\hat{c}_j|>10^{-5}$) gives TP=100\%, FP=0\%, and $F_1=1.000$
on an intentionally easy far-from-threshold family, with $F_1\ge 0.990$
across the full grid (Fig.~\ref{fig:supp_s1}c). Likewise, the Jacobi ablation
uses smooth controls only at $\alpha\in\{0.60,0.80\}$; over
$K\in\{2,4,6,8,10\}$ and $\beta\in\{0.30,0.45,0.55,0.70\}$ it maintains
TP=100\% and yields a narrow MAE plateau (0.065--0.080) on sharp profiles. We
therefore read panels (c)--(d) as sanity checks away from
$\alpha_{\mathrm{crit}}=0.4$, not as near-threshold calibration proofs.
(Fig.~\ref{fig:supp_s1}d).

\begin{table}[t]
\centering
\caption{Compact summary of the direct vorticity-threshold sweep on \texttt{isotropic1024coarse}. The same shared pool of 400 candidate centers and 120 requested profiles per threshold is used throughout.}
\label{tab:threshold_sweep}
\begin{tabular}{ccccc}
\hline
$|\omega|$ threshold & $n_{\mathrm{valid}}$ & $f_{\mathrm{req}}(\hat\alpha<0.4)$ & $\rho(\hat\alpha,|\omega|)$ & $p$ \\
\hline
$>3$  & 111 & 28.3\% & $-0.364$ & $8.5\times10^{-5}$ \\
$>5$  & 111 & 28.3\% & $-0.352$ & $1.5\times10^{-4}$ \\
$>8$  & 112 & 31.7\% & $-0.330$ & $3.8\times10^{-4}$ \\
$>10$ & 112 & 33.3\% & $-0.383$ & $3.0\times10^{-5}$ \\
$>12$ & 114 & 35.8\% & $-0.453$ & $4.3\times10^{-7}$ \\
\hline
\end{tabular}
\end{table}
\section{Physical Characterization}
\subsection{Dissipation Correlation}
\begin{figure*}[t]
  \centering
  \includegraphics[width=\textwidth]{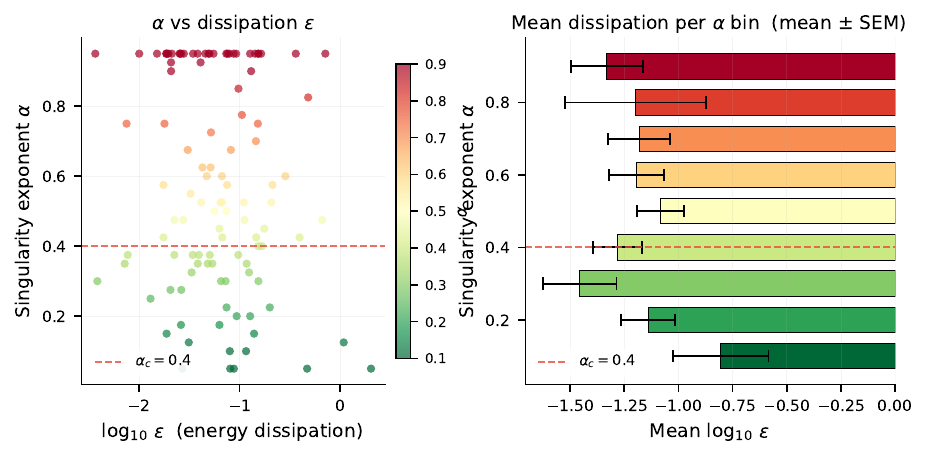}
  \caption{Dissipation--singularity correlation ($n{=}109$ JHTDB
    profiles). \emph{Left}: scatter of detected $\hat{\alpha}$ versus
    $\log_{10}\varepsilon$, colored by $\hat{\alpha}$. The plotted variables
    show only weak association, and the monotonic rank correlation is
    non-significant. For comparison, the Pearson correlation between
    $1/\hat{\alpha}$ and $\log_{10}\varepsilon$ is $r=0.235$ ($p=0.014$),
    explaining only 5.5\% of the variance.
    \emph{Right}: mean $\log_{10}\varepsilon$ per $\alpha$-bin ($\pm$SEM);
    extreme dissipation events ($\epsilon>\mu+2\sigma$) appear in both
    sharp and smooth profiles.\cite{jhtdb,meneveau1991}}
  \label{fig:dissipation}
\end{figure*}

A central physical prediction of multifractal intermittency theory is that intense energy dissipation $\epsilon$ is concentrated on a sparse set of small measure, corresponding to the low tail of the singularity spectrum $f(\alpha)$.\cite{meneveau1991,frisch1995,sreenivasan1997}
Our sparse-recovery method provides a conditional comparison of detected
roughness with local dissipation values queried from the JHTDB dataset.
At each of 120 high-vorticity centers, we compute the local dissipation rate
$\epsilon = 2\nu S_{ij} S_{ij}$ using the JHTDB \texttt{fd4lag4} finite-difference gradient operator, where $S_{ij} = \frac{1}{2} (\partial_i u_j + \partial_j u_i)$ is the strain-rate tensor.\cite{jhtdb}
We then correlate the detected effective exponent $\alpha$ (from $N = 40$ profiles) with $\log_{10}(\epsilon)$, yielding 109 valid pairs after filtering.
Figure~\ref{fig:dissipation} summarizes the dissipation--sharpness relationship.

Results show a weak and parametrization-dependent association. Pearson
correlation between $1/\hat{\alpha}$ and $\log_{10}(\epsilon)$ is $r = 0.235$
(95\% CI $[0.049, 0.405]$ via Fisher transformation; $p = 0.014$), but this
explains only $r^2 \approx 0.055$ of the variance. In contrast, the Spearman
rank correlation between $\hat{\alpha}$ and $\log_{10}(\epsilon)$ is
non-significant ($\rho = -0.128$, $p = 0.186$). The sign and significance
therefore depend on the chosen parametrization, and no simple monotonic law is
supported by the present sample.

Mean $\epsilon$ varies non-monotonically across $\alpha$ bins. The sharp-vs-smooth mean difference is non-significant ($p = 0.271$). These results caution against interpreting $\hat{\alpha}$ as a dissipation surrogate.

\subsection{Lagrangian singularity tracking}

To investigate whether the detected sharp singularities represent dynamically persistent material structures or transient Eulerian snapshots, we perform Lagrangian particle tracking in the JHTDB isotropic flow.
We seed $N=30$ fluid particles at high-vorticity centers identified at $t=0$, then advect them forward using first-order Euler integration:
\[
\mathbf{x}_{k+1} = \mathbf{x}_k + \mathbf{u}(\mathbf{x}_k, t_k) \Delta t,
\]
with stored-step increment $\Delta t_{\mathrm{store}} = 0.002$ in the JHTDB dataset parameterization.\cite{jhtdb} We do not convert this quantity into a Kolmogorov time or a laboratory-time interpretation here. At each of the subsequent $T=10$ stored time steps, we extract radial velocity increment profiles $f(r) = |\mathbf{u}(\mathbf{x}_k + r \hat{n}) - \mathbf{u}(\mathbf{x}_k)|$ centered at the current particle position (random direction $\hat{n}$), apply the $\alpha$ detector with $N=40$ samples, and compute three ensemble statistics: (i) the mean detected exponent $\overline{\alpha}(t)$; (ii) the sharp fraction $P(\alpha < 0.4 \mid t)$; and (iii) the temporal autocorrelation $C(\tau) = \langle \alpha(t) \alpha(t+\tau) \rangle / \langle \alpha(t)^2 \rangle$ over lags $\tau=0,\dots,5$. Because the direction is resampled at each step, these Lagrangian statistics conflate temporal evolution with directional variability; we therefore interpret the rapid decorrelation as an upper bound on persistence rather than a direction-controlled material invariant.

Results (Table~\ref{tab:lagrangian}) show short persistence under this
protocol. Mean $\hat{\alpha}$ rises from $0.565$ to $0.692$, and the sharp
fraction falls from $37.0$\% to $21.4$\%. The mean persistence of the binary
sharp label is $1.26$ stored steps. The
current saved artifact also yields a significant linear fit to the
$\overline{\alpha}(t)$ series, but given the short horizon, small ensemble, and
direction resampling, we treat that fit as descriptive only rather than as a
physically meaningful law of material evolution.

\begin{table}[ht]
\centering
\caption{Lagrangian evolution of detected singularities ($N=30$ particles, $T=10$ stored steps, $\Delta t_{\mathrm{store}}=0.002$ in the JHTDB dataset parameterization).}
\label{tab:lagrangian}
\begin{tabular}{c|ccc}
\hline
$t/\Delta t_{\mathrm{store}}$ & $\overline{\alpha}(t)$ & $P(\alpha<0.4)$ & $n_{\mathrm{det}}$ \\
\hline
0 & 0.565 & 0.370 & 27 \\
1 & 0.595 & 0.276 & 29 \\
2 & 0.541 & 0.333 & 27 \\
3 & 0.589 & 0.296 & 27 \\
4 & 0.555 & 0.308 & 26 \\
5 & 0.600 & 0.393 & 28 \\
6 & 0.580 & 0.286 & 28 \\
7 & 0.642 & 0.233 & 30 \\
8 & 0.623 & 0.233 & 30 \\
9 & 0.692 & 0.214 & 28 \\
\hline
\end{tabular}
\end{table}

These dynamics suggest that the detected sharp features are short-lived
\emph{Eulerian signatures} anchored to instantaneous high-vorticity points.
Given the random-direction sampling, this should be interpreted only as an
upper bound on persistence rather than as a direction-controlled material
invariant.\cite{meneveau1991,sreenivasan1997}
\begin{figure*}[t]
  \centering
  \includegraphics[width=\textwidth]{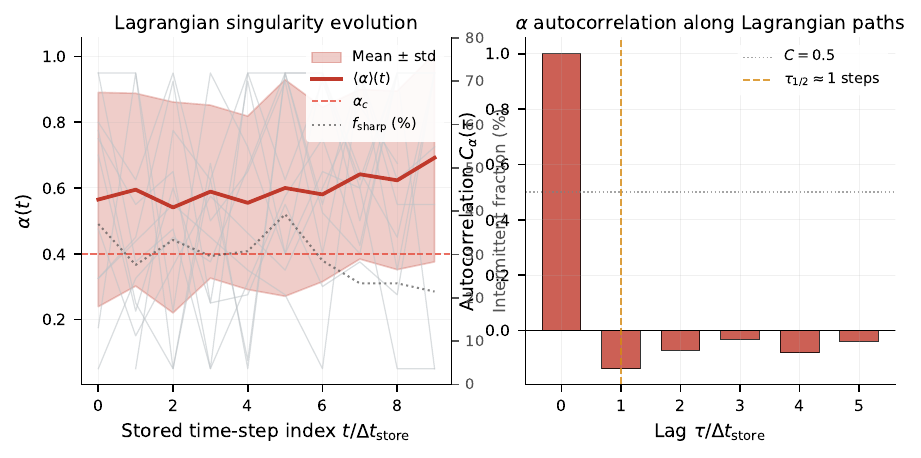}
  \caption{Lagrangian singularity evolution ($N{=}30$ particles,
    $T{=}10$ stored steps, $\Delta t_{\mathrm{store}}{=}0.002$ in the JHTDB
    dataset parameterization; mean persistence 1.26 steps).
    \emph{Left}: individual $\alpha(t)$ trajectories (gray) with
    ensemble mean\,$\pm$\,std (red band) and intermittent fraction
    $f_{\mathrm{sharp}}$ (dotted, right axis).
    \emph{Right}: autocorrelation $C_\alpha(\tau)$, showing rapid loss of
    correlation under the direction-resampled protocol.\cite{jhtdb,meneveau1991}}
  \label{fig:lagrangian}
\end{figure*}

\section{Reynolds Number Dependence}
\subsection{Intermittency across Reynolds numbers: scale effects and normalization}

Multifractal turbulence theory predicts that the inertial range broadens with increasing Reynolds number, potentially amplifying the prevalence of rare low-$\alpha$ singularities that dominate high-order structure functions.\cite{sreenivasan1997,meneveau1991,frisch1995}
To test this with local diagnostics, we apply the sparse M\"untz--Sz\'asz detector to high-vorticity profiles from three JHTDB isotropic datasets: \texttt{isotropic1024coarse} ($Re_\lambda \approx 433$), \texttt{isotropic4096} ($\approx 610$), and \texttt{isotropic8192} ($\approx 1300$).\cite{jhtdb}
For each dataset, we request $N=300$ profiles at randomly oriented directions
from high-vorticity centers selected with a fixed numerical vorticity cutoff in
the code, using $N=40$ samples per profile. We then compute the intermittent
fraction $f = \mathbb{P}(\alpha < 0.4)$ with bootstrap 95\% confidence
intervals (10000 resamples). These statistics are conditional on the
high-vorticity sampling rule and should be interpreted as
$\mathbb{P}(\alpha < \alpha_c \mid |\omega|>\text{threshold})$ rather than as
global intermittency fractions.

\paragraph{Fixed probe length (raw trend).} At fixed probe length
$r_{\max}=0.2$, the detected intermittent fraction is $37.8\%$
($Re_\lambda=433$), $41.1\%$ ($610$), and $48.3\%$ ($1300$), with mean
$\hat{\alpha}$ decreasing from $0.558$ to $0.488$. However, this trend is
confounded by the changing ratio of probe length to Kolmogorov scale:
$r_{\max}/\eta_K$ increases from $\sim 70$ to $\sim 294$ across the Reynolds
ladder, expanding the inertial range accessible to the detector.

\paragraph{Scale-normalized control.} With $r_{\max}/\eta_K \approx 70$ fixed, intermittency fractions are $34.4$\%, $27.8$\%, and $40.1$\% (bootstrap 95\% CIs: $[28.7\%, 39.8\%]$, $[22.7\%, 33.3\%]$, $[34.4\%, 45.7\%]$)---variable with overlapping CIs. The monotonic trend does not persist under scale normalization.

\paragraph{Interpretation.} Neither protocol establishes statistically significant Reynolds-number dependence or independence. The fixed-window increase may reflect measurement-window bias, while scale-normalized variability suggests genuine fluctuations or sampling limitations. We cannot claim ``universal'' $\sim 40$\% intermittency; the safest conclusion is that $30$\%--$50$\% of high-vorticity points exhibit rough local scaling, with no clear monotonic trend across the tested $Re_\lambda$ range.

\begin{table}[ht]
\centering
\caption{Intermittency fractions across Reynolds-number ladder at fixed physical $r_{\max}$ ($N=300$ profiles/dataset, bootstrap 95\% CIs).}
\label{tab:reynolds}
\begin{tabular}{lcccc}
\hline
Dataset & $Re_\lambda$ & $f(\alpha < 0.4)$ & 95\% CI & $\overline{\alpha}$ \\
\hline
isotropic1024coarse & 433 & 37.8\% & [32.1\%, 43.6\%] & 0.558 \\
isotropic4096 & 610 & 41.1\% & [35.4\%, 46.8\%] & 0.538 \\
isotropic8192 & 1300 & 48.3\% & [42.6\%, 54.0\%] & 0.488 \\
\hline
\end{tabular}
\end{table}
\begin{figure*}[t]
  \centering
  \includegraphics[width=\textwidth]{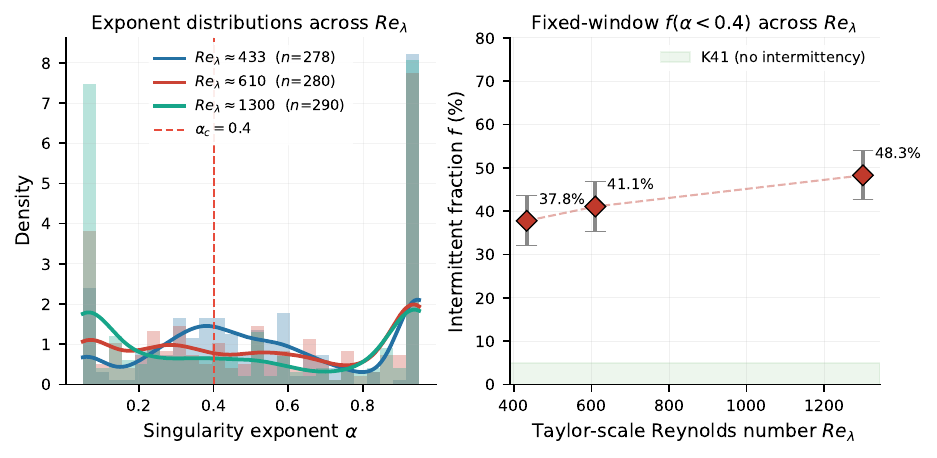}
\caption{Reynolds-number sweep at fixed physical $r_{\max}$.
    \emph{Left}: kernel-density estimates of the detected $\alpha$
    distribution for three JHTDB datasets ($n{=}278$--$290$ converged
    profiles each); higher $Re_\lambda$ shifts weight below
    $\alpha_c{=}0.4$ (dashed).
    \emph{Right}: intermittent fraction $f(\alpha{<}0.4)$ with
    bootstrap 95\% confidence intervals for the fixed-window sweep;
    the scale-normalized control in Fig.~\ref{fig:supp_s1}b removes
    any clear monotonic trend.\cite{jhtdb,sreenivasan1997}}
  \label{fig:reynolds}
\end{figure*}

These Reynolds-number results are qualitatively compatible with a persistent
low-$\alpha$ tail in high-vorticity regions, but they are not sufficient to
discriminate among phenomenological intermittency models. The dominant message
is methodological: the apparent fixed-window increase is entangled with the
widening physical scale window. Future work with matched band limits, unified
detector grids, and regenerated supplementary figures is needed before making a
strong Reynolds-number claim.

\section{Discussion}

\subsection{Comparison to full-field H\"older diagnostics}

Wavelet-leader and related full-field diagnostics estimate local regularity from
rich multiscale volumetric data.\cite{nguyen2019,muzy1991,arneodo2002,jaffard2016}
Our setting is different: we work with short, boundary-anchored one-dimensional
profiles. The present method is therefore complementary to full-field H\"older
mapping, not a replacement for it. In particular, this manuscript does not
provide a direct cross-calibration against wavelet-leader estimates on a common
dataset, so no quantitative equivalence claim should be inferred.

\subsection{Interpretation of low-$\alpha$ events}

The low-$\alpha$ events reported here are finite-band measurements on viscous
DNS data. They do \emph{not} establish pointwise singularities, loss of spatial
analyticity, or a direct connection to Onsager-critical regularity. At most,
they show that high-vorticity regions frequently generate steep local increment
profiles over the sampled radial band. Any bridge from these finite-resolution
measurements to mathematical singularity scenarios would require substantially
stronger analysis than is presently available.\cite{isett2018,hou2022}

\subsection{Geometric vs.~energetic observables}

Because the association between $\hat{\alpha}$ and dissipation is weak,
$\hat{\alpha}$ is not redundant with $\epsilon$ in the present sample. The weak
association suggests that the detected exponent captures aspects of the
geometric organization of short velocity-increment profiles that are not
exhausted by local energetic intensity. This empirical finding aligns with
recent geometrically oriented perspectives on turbulence, in which flow
organization is linked to dynamical response structure rather than to amplitude
observables alone.\cite{sevilla2026} The weak association with the local
invariant $Q$ and the more consistent negative association with vorticity
magnitude $|\omega|$ likewise suggest that the detected roughness may reflect
aspects of vortical organization not fully summarized by single pointwise
invariants, motivating future analysis within geometric-dynamics frameworks.
While the present diagnostic does not
resolve the full stress-propagator structure required to test that framework
directly, the observed separation between geometric and energetic observables
provides qualitative support for such nonlocal interpretations.

\paragraph{Directional controls.}
Auxiliary directional-geometry controls sharpen this interpretation while still
remaining short of a theorem-level validation. In a 60-center
high-vorticity-conditioned probe, the vorticity-aligned contrast
$\Pi_\alpha=\hat{\alpha}_{\parallel}-\hat{\alpha}_{\perp}$ is positive on
average ($\langle \Pi_\alpha \rangle \approx 0.093$, bootstrap 95\% CI
$[0.028,0.158]$, one-sample $p<10^{-2}$). The same probe gives weaker or
vanishing contrast on shuffled or random fake axes, and the true-vorticity axis
outperforms the compressive and extensional strain eigenframes. A companion
random-direction angular fit over 32 directions per center yields a positive
mean $P_2(\cos\theta)$ slope and a joint Legendre decomposition with significant
mean $c_2$ but non-significant mean $c_4$ and $c_6$; the normalized $l=2$
energy exceeds the corresponding $l=4$ and $l=6$ contributions in this
conditional sample. The prominence of the $l=2$ contribution is qualitatively
consistent with Sevilla's low-order anisotropic-sector hierarchy,\cite{sevilla2026}
and, taken together, these controls indicate that the observable contains a
reproducible low-order directional signature tied to vortical geometry. More
specifically, $\Pi_\alpha$ is useful because it compresses the
short-profile directional response into a signed scalar with a clear geometric
reference axis and a falsifiable null hypothesis: if the recovered roughness
were insensitive to local geometric organization, no systematic difference
between the parallel and transverse estimates would be expected. In that sense,
the present paper does more than report another correlation with $|\omega|$:
it introduces a concrete finite-scale observable that can be re-used, modeled,
or challenged in later theory-facing work.

\paragraph{Scale-transfer scans.}
The updated seeded scale-transfer scan over
$r_{\max}\in\{0.06,0.20,0.57\}$ on 60 high-vorticity centers refines that
picture. The vorticity-aligned contrast $\Pi_\alpha$ is positive and
statistically significant at both the smallest and largest radii tested, while
the intermediate radius gives a weaker, non-significant mean. The amplitude is
therefore non-monotone across scales, but the large-minus-small difference is
itself non-significant, so the present data do not show a clear collapse of the
axis-specific signal at larger probe radius. At the same time, the broader
anisotropy summaries $\Delta_\alpha$ and $\mathrm{Var}(\hat{\alpha})$ increase
with $r_{\max}$. We interpret these scans as evidence for finite-range
persistence of the directional signature rather than as a clean
scale-persistent or strongly nonlocal confirmation of a geometric-dynamics
theorem. At the same time, the pooled explained variance remains small, the
controls are conditioned on high-vorticity centers, and the present experiments
do not establish a full geometric-dynamics theorem or a dissipation-matched
separation principle.

\subsection{Extensions and outlook}

Several extensions are natural. The most urgent are methodological: unify the
alpha grid across experiments, extend the balanced synthetic control to the
same baseline-comparison suite and to more physically realistic profile
families, incorporate short-record ESS/local-slope comparators, and carry out
a full revalidation of quadrature-weighted detectors for the present
boundary-anchored sampling geometry.
On the application side, sparse radial profiling could be tested on experimental
transects, anisotropic flows, wall-bounded turbulence, or joint multi-direction
measurements that reduce directional ambiguity.

\subsection{Comparison with Existing Intermittency Diagnostics}

Our sparse-recovery roughness diagnostic is complementary to, rather than a
replacement for, established intermittency diagnostics:

\begin{itemize}
    \item \textbf{WTMM and Wavelet Leaders \cite{muzy1991,arneodo2002,jaffard2016}:} These methods estimate singularity spectra or local regularity from multiscale wavelet information. They are powerful when many octaves and interior singularities are available, but they are not designed for the short boundary-anchored profiles studied here.
    \item \textbf{Structure functions \cite{frisch1995,wang1996}:} These estimate anomalous moment scaling exponents $\zeta_p$ from ensemble averages. They characterize intermittency statistically, but they do not provide an event-level sharp/smooth decision on short individual profiles.
    \item \textbf{ESS and local slopes \cite{benzi1993}:} These methods estimate local scaling from short-range regressions. They are sensible comparison points for 1D data and should be included in stronger future baselines.
    \item \textbf{Flatness and hyperflatness \cite{sreenivasan1997}:} These summarize non-Gaussianity across scales, whereas the present detector returns a local model-based estimate on individual sampled profiles.
\end{itemize}

\subsection{Physical Implications of the Reynolds Number Sweep}

The scale-normalized control ($r_{\max}/\eta_K=70$) yields intermittent
fractions of $34.4\%$, $27.8\%$, and $40.1\%$ with fully overlapping bootstrap
confidence intervals. The fixed physical window $r_{\max}=0.2$ yields higher
fractions, rising from $37.8\%$ to $48.3\%$, but that increase is entangled
with the widening range $r_{\max}/\eta_K \approx 70 \to 294$. The safest
interpretation is therefore modest: the conditional high-vorticity sample shows
a substantial low-$\alpha$ tail across the Reynolds ladder, but the current
experiments do not isolate a clean intrinsic Reynolds-number law.

\subsection{Limitations and caveats}

We emphasize the following limitations:

\begin{enumerate}
    \item \textbf{Theory-to-code gap:} The geometric discussion is motivated by \cite{eng2026}, but the reported detector uses an unweighted discrete Lasso and does not verify the assumptions of the cited theory. The new discrete-dictionary diagnostic confirms that the implemented design matrices are also ultra-coherent and nearly rank-degenerate; it does \emph{not} close the theorem-to-implementation gap. A preliminary quadrature-weighted variant was explored, but it materially altered the empirical operating point under the present JHTDB protocol and is therefore not adopted as the primary method in this paper.
    \item \textbf{Inconsistent detector grids:} Different experiments use different $\alpha$ grids (37-point and 21-point versions), so the current manuscript cannot honestly describe all results as outputs of a single fixed detector.
    \item \textbf{Amplitude scaling:} The implementation does not normalize amplitudes by $u_{\mathrm{rms}}$, so the thresholds $\sigma=0.01$ and $|\hat{c}_j|>10^{-5}$ are empirical rather than scale-invariant.
    \item \textbf{Threshold-dependent summaries:} The binary sharp fraction depends materially on both the vorticity-conditioning threshold and the classification threshold $\alpha_{\mathrm{crit}}$, as quantified by the dedicated sensitivity sweeps. These summaries are therefore operating-point descriptors, not detector-independent population constants.
    \item \textbf{Validation gaps:} The current fBm stress test remains too biased and noisy to serve as a quantitative calibration result, and the JHTDB subsampling self-consistency benchmark is run-dependent: across nine saved unweighted reruns, the $N{=}40$ $F_1$ values span $0.84$ to $1.00$ with mean $0.930$ rather than defining a single fixed external-accuracy number.
    \item \textbf{Synthetic benchmark scope:} A balanced Lasso-based synthetic control is now available, but it still uses a simplified profile family and the multi-method baseline comparison remains restricted to sharp synthetic profiles only.
    \item \textbf{Dissipation sample size:} The dissipation analysis uses only $n=109$ valid, high-vorticity-conditioned pairs, and its qualitative conclusion is limited to the absence of a strong simple relationship.
    \item \textbf{Lagrangian protocol:} Direction resampling at each time step mixes temporal evolution with directional variability, so the persistence estimates are upper bounds rather than clean material diagnostics.
    \item \textbf{Directional-geometry controls:} The new vorticity-aligned and Legendre-mode controls are auxiliary, high-vorticity-conditioned probes. They support a weak but reproducible low-order directional signal, but the pooled explained variance is small and the present dissipation-matched comparisons are not yet strong enough to justify a clean geometry-versus-energy separation claim.
    \item \textbf{Scale-transfer scans:} The seeded multi-$r_{\max}$ scan now shows positive $\Pi_\alpha$ at both the smallest and largest tested radii, with a non-significant large-minus-small difference, while broader anisotropy measures increase with $r_{\max}$. This supports finite-range persistence of the directional signal, but the scale dependence remains non-monotone and still falls short of a strong scale-persistent or nonlocal theorem-level claim.
\end{enumerate}

% ============================================================
\section{Conclusion}
% ============================================================

We have presented a sparse-recovery study of finite-scale roughness in 3D
turbulence using short radial profiles. The strongest numerical result at
$N=40$ is an internal JHTDB subsampling benchmark, where across nine saved
unweighted reruns the sharp/smooth classifier spans $F_1 \in [0.84,1.00]$ with mean
$0.930$ relative to $N=200$ labels produced by the same detector, and a
recent unweighted run gives $F_1 = 0.927$. A balanced synthetic Lasso control
gives balanced accuracy $0.928$ at $N=40$ with perfect sharp-class recall but
specificity $0.855$, indicating useful numerical sanity with nontrivial smooth
false positives rather than a complete external calibration.

Across the JHTDB Reynolds ladder, the conditional sharp fraction remains of
order $30$\%--$50$\% in high-vorticity regions, with strong dependence on the
chosen probing window. Dissipation is only weakly associated with the detected
roughness, and a short Lagrangian protocol shows rapid loss of persistence
under direction resampling. Taken together, these results support a cautious
interpretation of $\hat{\alpha}$ as a model-dependent finite-band roughness
indicator for sparse local measurements and, more specifically, as a geometric
observable of short boundary-anchored profiles that is only weakly coupled to
dissipation. They do not justify stronger claims about universal Reynolds
scaling, direct dissipation surrogacy, or mathematical singular behavior.
Auxiliary directional controls strengthen that cautious geometric reading: in a
60-center high-vorticity-conditioned probe, $\Pi_\alpha$ is significantly
positive, the true vorticity axis outperforms shuffled and random fake axes,
and a joint Legendre fit detects a weak but statistically stable low-order
quadrupolar component with $l=2$ dominating $l=4$ and $l=6$ in normalized
energy. A seeded scale-transfer scan over $r_{\max}=0.06,0.20,0.57$ further
shows that $\Pi_\alpha$ remains positive and statistically significant at the
smallest and largest tested radii, with no significant large-minus-small drop,
even though the amplitude varies non-monotonically across scales and the
broader anisotropy measures grow with $r_{\max}$. We therefore regard the
present evidence as supporting a finite-scale directional geometry effect with
finite-range persistence, and, more specifically, as elevating $\Pi_\alpha$ to
a practical observable for future geometry-based turbulence studies. We do not
regard the present evidence as validating a full Sevilla-style
geometric-dynamics theorem, as establishing a strong scale-persistent nonlocal
signature, or as establishing dissipation-matched independence.

\section*{Declaration of AI Tool Usage}
The authors used Kimi v2, ChatGPT 5.2, and Claude Opus 4.5 for code assistance. All content was critically reviewed and edited by the authors. The authors take full responsibility for the content.

\section*{Acknowledgments}
The authors acknowledge the Johns Hopkins Turbulence Database for providing open access to DNS datasets. We thank Dr.~Tianyi Li (University of Rome Tor Vergata) for suggesting the finite-scale diagnostic interpretation and validation on fractional Brownian motion, and Dr.~Alejandro Sevilla (Universidad Carlos III de Madrid) for extensive feedback on the geometric dynamics framework and the nonlocal propagator interpretation of the results.

\end{document}